\documentclass[letterpaper]{article}
\usepackage{aaai23} 
\usepackage{times} 
\usepackage{helvet} 
\usepackage{courier} 
\usepackage[hyphens]{url} 
\usepackage{graphicx} 
\usepackage{graphicx} 
\usepackage{natbib} 
\usepackage{caption} 
\usepackage{tikz}
\usepackage{cite}
\usepackage{soul}
\usepackage{xcolor}
\usepackage{subfig}
\usepackage{amsmath}
\usepackage{physics}
\usepackage{booktabs}
\usepackage{amsfonts}
\usepackage{textcomp}
\usepackage{algorithmic}
\usepackage{filecontents}
\usepackage[normalem]{ulem}
\usepackage[linesnumbered,lined,boxed,commentsnumbered]{algorithm2e}

\usepackage{numprint}
\npthousandsep{,}

\usepackage{xcolor,soul}

\usepackage{tcolorbox}
\newtcolorbox{highlighted}{colback=yellow,coltext=black,breakable}

\begin{document}

\title{\textsc{SliQ}: Quantum Image Similarity Networks on Noisy Quantum Computers}

\author {
    Daniel Silver,
    Tirthak Patel,
    Aditya Ranjan,
    Harshitta Gandhi,
    William Cutler,
    Devesh Tiwari
}
\affiliations {
  {Northeastern University}\\
  {\{silver.da, patel.ti, ranjan.ad, gandhi.ha, 
cutler.wi, d.tiwari\}@northeastern.edu}
}

\maketitle

\begin{abstract}

Exploration into quantum machine learning has grown tremendously in recent years due to the ability of quantum computers to speed up classical programs. However, these efforts have yet to solve unsupervised similarity detection tasks due to the challenge of porting them to run on quantum computers. To overcome this challenge, we propose \textsc{SliQ}, the first open-sourced work for resource-efficient quantum similarity detection networks, built with practical and effective quantum learning and variance-reducing algorithms. 

\end{abstract}

\section{Introduction}
\label{sec:intro}

\textbf{Brief Overview and Motivation.} Rapid advancements in quantum machine learning (QML) have increasingly allowed researchers to leverage the benefits of quantum computing in solving ML problems. Although the degree of advantage for different ML tasks is being explored, the recent advances suggest that classification and solving high energy physics problems are among the most promising candidates \cite{49725, guan2021quantum}. In particular, recent efforts have focused on developing high-quality classification circuits for quantum computers. While the resulting classifiers have been highly effective, they are restricted because classification, like all other supervised learning methods, requires labeled data. In many real-world scenarios, labeled data is either not readily available (e.g., diagnosing complex diseases based on medical images without quantifiable ground truth \cite{utkin2019ensemble}) or not feasible (e.g., a visual sketch of a suspect) \cite{wan2019transfer}. In such scenarios, comparison across unlabeled inputs is critical for learning, prediction, and ground truth generation -- hence the popularity of similarity detection for various tasks including recommendation systems \cite{wei2019personalized}. However, there is no QML circuit designed to predict similarity on unlabeled data currently available.   

\noindent\textbf{\textsc{SliQ}: Solution and Approach.} To bridge this gap, a na\"ive design for similarity detection might create pairs of similar and dissimilar inputs from the training dataset, much like classical Siamese and Triplet networks~\cite{Chicco2021,10.1007/978-3-030-58621-8_9} (e.g., Anchor-Positive and Anchor-Negative pairs), and pass them sequentially over a variational quantum circuit (VQC)\cite{cerezo2021variational} to minimize the loss function. While straightforward, this design is resource-inefficient and does not fully leverage the unique properties of quantum computing. To address this gap in unsupervised QML, we propose \textsc{SliQ}, which mitigates these challenges with multiple novel key design elements. 

First, \textsc{SliQ} addresses resource inefficiency by training both images in the pair at once via a superimposed state. This provides multiple advantages: (1) it reduces the overall quantum resource requirements by reducing the number of runs, and (2) it allows the VQC to learn more effectively since the superposition provides explicit hints about the similarity between the data. Next, to take advantage of entanglement in quantum computing systems, \textsc{SliQ} interweaves the features of both inputs and explicitly entangles them at each layer in the VQC, decreasing the distance of corresponding features in Hilbert space. \textsc{SliQ}'s design ensures that interwoven features from different inputs are embedded into all physical qubits of the learning circuit so that the entanglement effects are captured in the measurement qubits. To ensure that \textsc{SliQ} is practical and effective on current error-prone quantum computers, \textsc{SliQ} keeps the number of parameters in the learning circuit minimal to mitigate the compounding noise effects on real quantum computers.

Unfortunately, incorporating superposition and entanglement properties creates new challenges. The identities of individual inputs in a pair (e.g., Anchor input in the Anchor-Positive pair) are indistinguishable due to entanglement, and the projection of the same input on the classical space is inconsistent across different runs. \textsc{SliQ} introduces a new training ``loss'' estimation and improves quantum embedding methods to reduce projection variance, resulting in more robust training and a network that is resilient to hardware errors on real quantum systems. Overall, \textsc{SliQ} demonstrates that the combination of training on entangled pairs and utilizing a projection variance-aware loss estimation yields effective similarity detection, even on current noisy quantum computers. 

\noindent\textbf{Contributions of \textsc{SliQ}.} 

\noindent \textbf{I.} To the best of our knowledge, \textsc{SliQ} is the first method to build a practical and effective \textit{quantum learning circuit for similarity detection on NISQ-era quantum computers}. \textsc{SliQ} is available as open-source framework at \texttt{https://github.com/SilverEngineered/SliQ}.

\noindent \textbf{II.} \textsc{SliQ}'s design demonstrates how to exploit the superposition and entanglement properties of quantum computing systems for similarity detection. It builds a \textit{resource-efficient training pipeline by creating interwoven, entangled input pairs on a VQC, and it applies new robust methods for the quantum embedding of classical inputs.} 

\noindent \textbf{III.} Our simulations and real-computer evaluations demonstrate that \textsc{SliQ} achieves a 31\% point improvement in similarity detection over a baseline quantum triplet network on a real-world, unlabeled dataset~\cite{chen2018cartoongan}, while prior state-of-the-art works in QML only perform classification and require labeled input data~\cite{49725,silver2022quilt}. We also show that \textsc{SliQ} performs competitively for classification tasks on labeled data, despite not being a primary objective a similarity network. 

\section{Background}

\noindent\textbf{Qubits, Quantum Gates, and Quantum Circuits.} A quantum bit (qubit) has the ability to attain a \textit{superposition} of its two basis states: $\ket{\Psi} = \alpha_0\ket{0} + \alpha_1\ket{1}$. Here, $\ket{0}$ and $\ket{1}$ are the two basis states, $\alpha_0$, and $\alpha_1$ are normalized, complex-valued coefficients and $\Psi$ is the overall qubit state in superposition. For an $n$-qubit computing system, the overall system state is represented as: $\ket{\Psi} = \sum_{k=0}^{k=2^n-1}\alpha_k\ket{k}$. When this state is \textit{measured}, the \textit{superposition collapses}, and the system is observed in state $\ket{k}$ with probability $\norm{\alpha_k}^2$.

A qubit can be put in arbitrary superpositions using the $R3(p_1, p_2, p_3)$ \textit{quantum gate}. The single-qubit $R3$ gate has three parameters ($p_1, p_2, p_3$) that can be adjusted to achieve the desired state~\cite{cerezo2021variational}. Multiple qubits can be \textit{entangled} together to form an $n$-qubit system using two-qubit gates (e.g., the $CX$ gate). These non-parameterized two-qubit gates and tunable $R3$ gates may be combined to achieve any $n$-qubit computation.

\begin{figure}[t]
    \centering
    \includegraphics[scale=0.52]{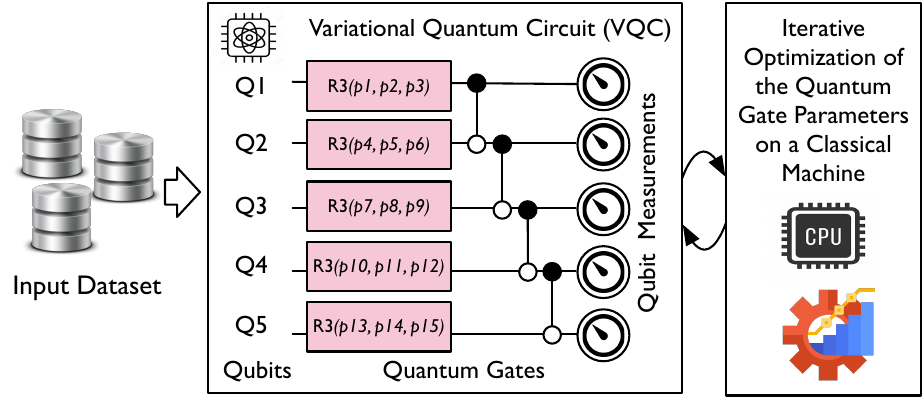}
    \caption{Hybrid quantum-classical procedure of executing and optimizing a variational quantum circuit (VQC).}
    \label{fig:varia}
\end{figure}

A sequence of quantum gates applied to a system of qubits forms a \textit{quantum circuit}, at the end of which the qubits are measured to obtain the circuit's output. Fig.~\ref{fig:varia} shows an example of a quantum circuit in the ``Variational Quantum Circuit (VQC)'' box. The horizontal lines represent five qubit states, to which the $R3$ and two-qubit gates are applied over time, and the measurement gates are applied at the end.

\noindent\textbf{Variational Quantum Circuits and Quantum Machine Learning.} Whereas the gates in a normal quantum circuit are deterministic and predefined, a \textit{Variational Quantum Circuit (VQC)} is a quantum circuit that utilizes parameterized gates that are tuned to optimize a certain objective~\cite{cerezo2021variational}. This objective can take many forms, from finding the minimal energy of a molecule's Hamiltonian to maximizing the rate of return of a financial portfolio or, in \textsc{SliQ}'s case, optimizing the loss function of a \textit{quantum machine learning (QML)} task~\cite{nisq,Aaronson2015}. These circuits are ``variational'' because the gates' parameter values vary during the optimization process. These gates are adjusted according to a classical optimizer running on a classical machine, while the variational circuit itself is executed on a quantum computer. Fig.~\ref{fig:varia} demonstrates this hybrid feedback approach between quantum execution and classical optimization.

Although the optimization of VQC parameters is performed classically, a quantum advantage can be obtained from the circuit's execution on quantum hardware, which means the circuit has far fewer parameters to optimize compared to the classical version of the same algorithm~\cite{chen2020variational}. This advantage is gained by utilizing the superposition and entanglement properties on quantum computers that are not available on classical computers. For example, a classical image classification neural network typically consists of millions of parameters while a well-designed quantum network only requires hundreds~\cite{cerezo2021variational}. However, the accuracy of such quantum networks has been limited due to prevalent noise on current quantum computers~\cite{nisq}, especially for unsupervised learning~\cite{cerezo2021variational} -- this paper aims to overcome this barrier.

\noindent\textbf{Noise on NISQ Computers.} Contemporary quantum computers suffer from noise during program execution due to various imperfections in the hardware of physical qubits, causing errors in the program output. \textsc{SliQ} aims to achieve effective results in the face of these challenges.

\section{\textsc{SliQ}: Design and Implementation}
\label{sec:solution}

\begin{figure*}[t]
    \centering
    \includegraphics[scale=0.6]{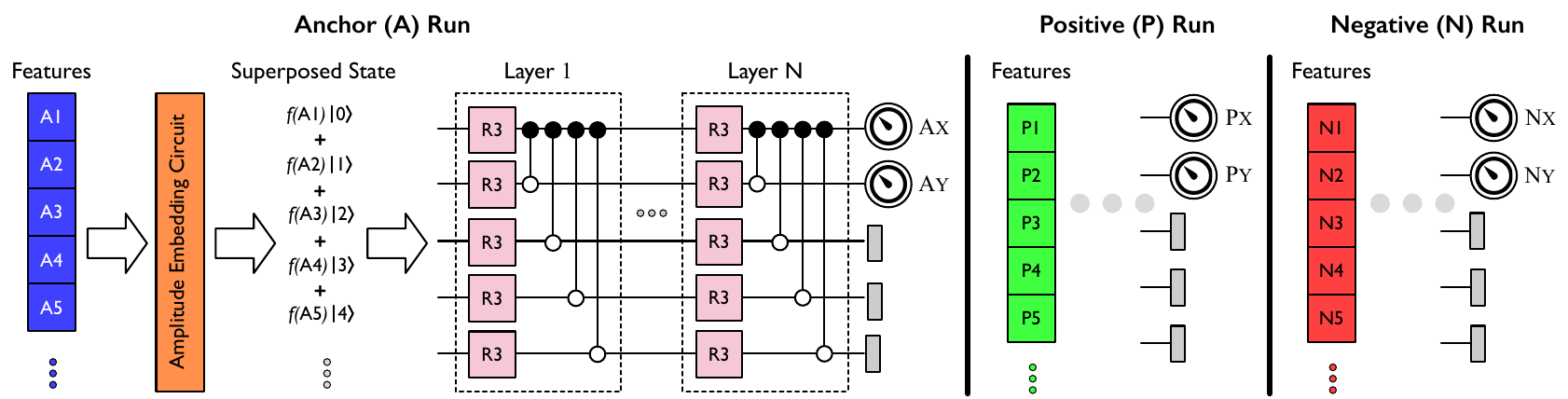}
    \caption{The baseline design inputs one image at a time, requiring three separate runs for A, P, and N.}
    \label{fig:baseline}
\end{figure*}

\begin{figure*}[t]
    \centering
    \includegraphics[scale=0.6]{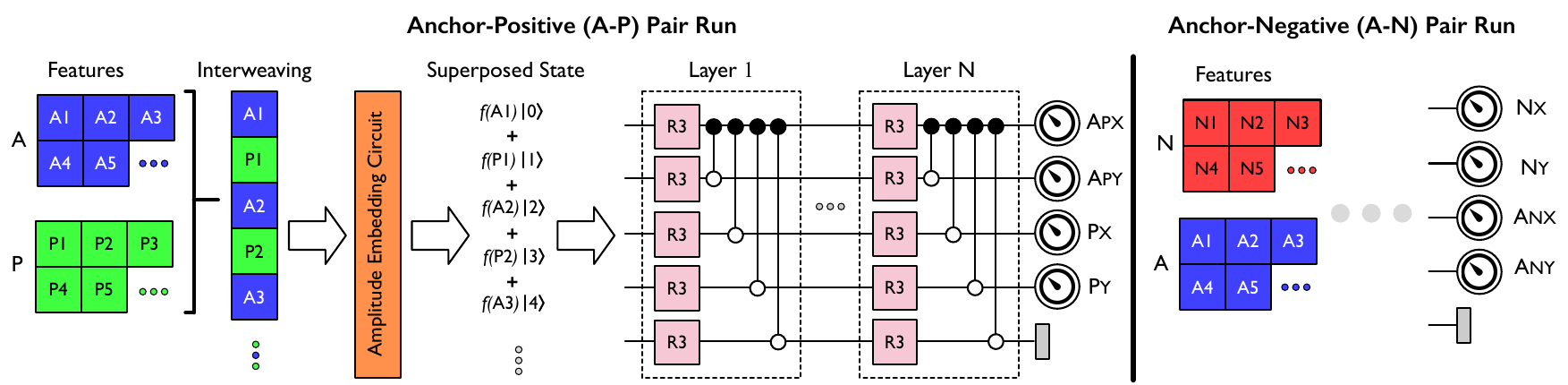}
    \caption{\textsc{SliQ}'s procedure of combining A-P and A-N inputs into two runs, interweaving their feature space, and updating the variational structure to leverage the properties of quantum superposition and entanglement to reduce the number of runs and generate better results.}
    \label{fig:key_idea_1}
\end{figure*}

In this section, we discuss the design of \textsc{SliQ} and its key design elements. Before we discuss the design details of \textsc{SliQ}, we first describe a base design of a quantum machine learning circuit for similarity detection. We refer to this design as \textit{``Baseline''}. First, we discuss how the baseline design leverages widely-used variational quantum circuits to build a quantum learning network to perform the similarity detection task for labeled/unlabeled data. Next, we discuss the baseline design's resource inefficiency and its inability to exploit the power of superposition and entanglement. \textsc{SliQ}'s design addresses those limitations and provides superior performance, as discussed in our evaluation.


\noindent\textbf{\newline Baseline Design.} The baseline design has three major components. The first component is generating the input which consists of triplets of Anchor, Positive, and Negative inputs -- similar to classical Siamese-based and Triplet models which are widely used in the classical similarity networks~\cite{veit2017conditional, ma2020fine}. Then the encoding of the input features to the physical qubits is performed. To achieve this, we perform amplitude embedding \cite{schuld2018supervised} on the inputs one by one for all three inputs (Fig.~\ref{fig:baseline}). The amplitude embedding procedure embeds classical data as quantum data in a Hilbert Space. Although recent prior works have utilized principal component analysis (PCA) prior to amplitude embedding for feature encoding \cite{silver2022quilt}, the baseline does employ PCA because higher performance is observed with keeping features intact and padding 0's as necessary to maintain the full variance of features. The second component is to feed these encoded features to a VQC for training and optimizing the VQC parameters to minimize the training loss (Fig.~\ref{fig:baseline}). The training loss is estimated via calculating the distance between the projection of inputs on a 2-D space -- the projection is obtained via measuring two qubits at the end of the VQC circuit (this is the third and final component). The loss is calculated as the squared distance between the projection of the anchor, $A$, to the positive projection, $P$, taking the absolute distance between the anchor projection and negative projection $N$. This is an $L2$ variant of Triplet Embedding Loss and is formally defined as 
{\small
\begin{equation}
L_2 = \Big{\{}(A_x- P_x)^2 + (A_y-P_y)^2\Big{\}} - \Big{\{}(A_x- N_x)^2 +(A_y- N_y)^2\Big{\}}
\end{equation}
}

We note that the effectiveness of training can be adapted to any choice of dimension and shape of the projection space (2-D square box bounded between -1 and 1, in our case) as long as the choice is consistent among all input types (Anchor, Positive, and Negative inputs). A more critical feature is the repeating layers of the VQC which the baseline design chooses to be the same as other widely-used VQCs to make it competitive \cite{Chicco2021,10.1007/978-3-030-58621-8_9}.

\subsection{\textsc{SliQ}: Key Design Elements}

\textsc{SliQ} builds off the baseline design and introduces multiple novel design aspects to mitigate the limitations of the baseline design. First, \textsc{SliQ} introduces the concept of training an input pair in the same run to leverage the superposition and entanglement properties of quantum computing systems.

\noindent\textbf{Input Feature Entanglement and Interweaving.}  Recall that in the baseline design, each image type (Anchor, Positive, Negative) traverses through the variational quantum circuit one-by-one. The corresponding measurements at the end of each image-run produce coordinates on a 2-D plane that allows us to calculate similarity distance between A-P and A-N inputs. This allows us to calculate the loss that is targeted to be minimized over multiple runs during training. Unfortunately, this procedure requires performing three runs before the loss for a single (A, P, N) triplet input can be estimated, which is resource inefficient. 

\textsc{SliQ}'s design introduces multiple new elements to address this resource inefficiency. The first idea is to create two training pairs (Anchor-Positive and Anchor-Negative), and each pair is trained in a single run (demonstrated visually in Fig.~\ref{fig:key_idea_1}). This design element improves the resource-efficiency of the training process -- instead of three runs (one for each image type), \textsc{SliQ} requires only two runs. Note that, theoretically, it is possible to combine all three input types and perform combined amplitude embedding. However, in practice, this process is not effective because it becomes challenging for the quantum network to learn the distinction between the positive and negative input relative to the anchor input. Creating two pairs provides an opportunity for the quantum circuit to learn the similarity and dissimilarity in different pairs without dilution. 

The second idea is to interweave the two input types in a pair before performing the amplitude embedding, and then feeding the output of the amplitude embedding circuit to the quantum circuit (Fig.~\ref{fig:key_idea_1}). Interweaving provides the advantage of mapping features from different inputs to different physical qubits. This is particularly significant to mitigate the side-effects of noise on the current NISQ-era quantum machines where different physical qubits suffer from different noise levels~\cite{tannu2019not}. If interweaving of images is not performed, we risk the network not learning direct comparison between positionally equivalent features. \textsc{SliQ}'s interweaving mitigates this risk to make it effective on NISQ-era quantum computers which we found when compared to layering images.


As a final remark, we note that all these ideas are combined to leverage the power of entanglement and superposition of quantum systems -- by training multiple inputs together, interweaving them, and creating superposition, and entangling them. While \textsc{SliQ}'s design to exploit superposition and entanglement is useful, it creates new challenges too. Next, we discuss the challenges of attaining projection invariance and novel solutions to mitigate the challenge.

\begin{figure}[t]
    \centering
    \includegraphics[scale=0.6]{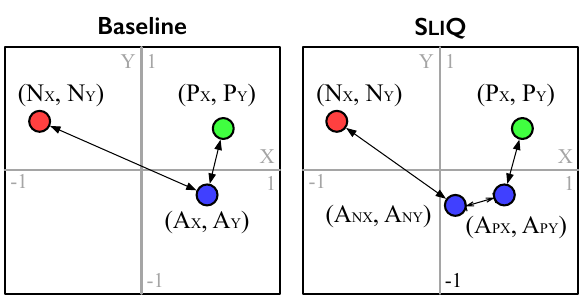}
    \caption{While the baseline design has zero projection variance, \textsc{SliQ} has to take additional steps to mitigate it.}
    \label{fig:key_idea_2}
\end{figure}

\noindent\textbf{Projection Variance Mitigation (PVM).}  Recall that in the baseline design, we measure two qubits and project the input's position in a 2-D space. Over three runs, we receive three separate coordinates in 2-D space, which we can use to calculate the loss -- as shown in Fig.~\ref{fig:key_idea_2} (left). Our objective is to minimize the overall loss, defined as below:
{\small
\begin{equation}
L_{obj} = \Big{(}\abs{A_x- P_x} + \abs{A_y- P_y}\Big{)} - \Big{(}\abs{A_x- N_x} + \abs{A_y- N_y}\Big{)}
\end{equation}
}

Optimizing for the above objective function is relatively straightforward. However, this loss function becomes non-trivial when \textsc{SliQ} introduces the idea of training input pairs. Recall that the inputs are interweaved (Fig.~\ref{fig:key_idea_1}), and hence, our measurements need to capture the outputs of anchor and positive/negative features separately. \textsc{SliQ} resolves these issues by increasing the number of qubits we measure. Instead of two qubits per run, \textsc{SliQ} measures four qubits. In Fig.~\ref{fig:key_idea_1}, these qubits are denoted at the end of both runs. To distinguish the anchor input, \textsc{SliQ} enforces two apriori-designated qubits measurements to correspond to the anchor in both runs. We note it is not critical which qubits are chosen to ``represent'' the anchor input as long as our choice is treated as consistent. For example, qubits 1 and 2 could be tied to the anchor image, or qubits 1 and 3. So long as the choice does not change through training and evaluation, these options are computationally identical. However, this idea creates a major challenge -- the coordinates corresponding to the anchor input may not project to the same point in our 2-D space. This is visually represented by two points $(A_{NX}, A_{NY})$ and $(A_{PX}, A_{PY})$ in Fig.~\ref{fig:key_idea_2}. Ideally, these two points should project on the same coordinates. 

The baseline design inherently has zero projection variance because it only has one measurement corresponding to the anchor input, and the loss for the positive and negative input was calculated from this absolute pivot. To mitigate this challenge, \textsc{SliQ} designs a new loss function that accounts for minimizing this projection variance over training. As shown below, \textsc{SliQ}'s novel loss function has two components: (1) the traditional loss between the positive/negative input and the anchor input, and (2) new consistency loss. Consistency loss enforces positional embeddings to separate the entangled images at the time of measurement.
\begin{equation}
\label{eq:l_p}
L_{pvm} = \abs{A_{px} - A_{nx}} + \abs{A_{py} - A_{ny}}
\end{equation}
\begin{equation}
\label{eq:loss}
L_{total} = \alpha * L_{obj} + \beta * L_{pvm}
\end{equation}

\begin{figure}[t]
    \centering
    \includegraphics[scale=0.55]{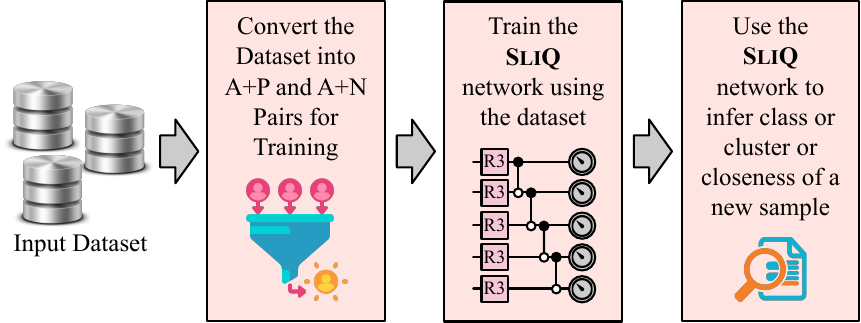}
    \caption{Overview of the design of \textsc{SliQ} including the steps of preparing the data for training, training the quantum network, and using the network for inference post-training.}
    \label{fig:sliq}
\end{figure}

\begin{figure*}[t]
    \centering
    \subfloat[Baseline]{\includegraphics[scale=0.44]{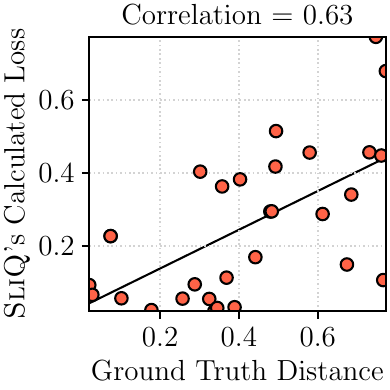}\hfill
    \includegraphics[scale=0.44]{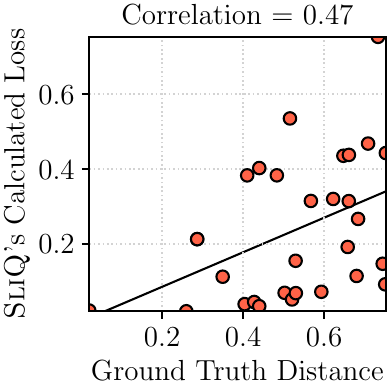}\hfill
    \includegraphics[scale=0.44]{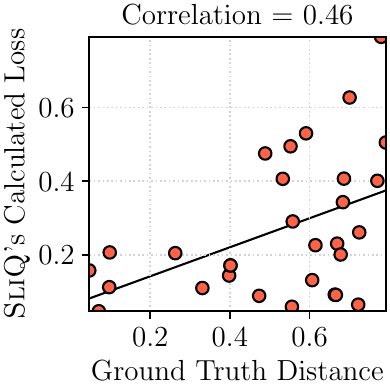}}\hfill
    \subfloat[\textsc{SliQ}]{\includegraphics[scale=0.44]{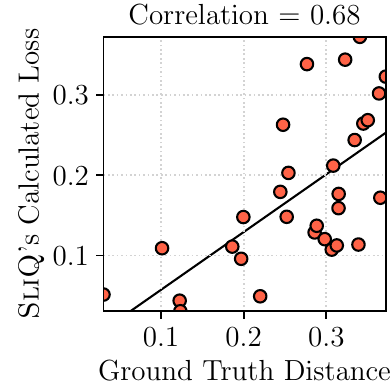}\hfill
    \includegraphics[scale=0.44]{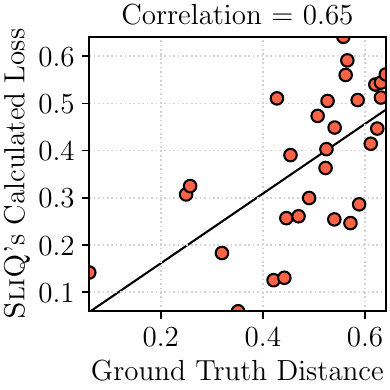}\hfill
    \includegraphics[scale=0.44]{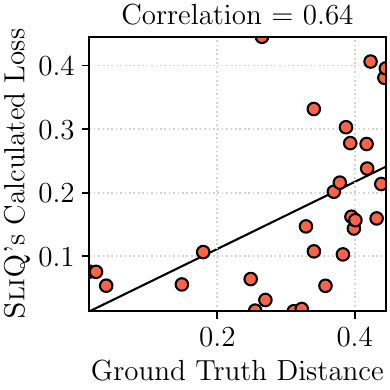}}
    \caption{\textsc{SliQ}'s loss normalized against ground truth distance between color frequencies. The anchor image was compared to 100 images. \textit{Only the top 3 correlations are shown.} The steep drop in correlation metric shows that the baseline is not effective.}
    \label{ranking} 
\end{figure*}

In Eq.~\ref{eq:loss}, the parameters $\alpha$ and $\beta$ are hyperparameters that denote weights for the objective function to balance the objective of accurate embeddings and precise embeddings. For $L_{obj}$, we use $(A_{px}, A_{py})$ and ($A_{nx}, A_{ny})$ for the positive and negative anchor values respectively.  Additionally, to ensure robustness across the circuit, the samples are reversed for the negative case.  The pair (A,P) is run along with the pair (N,A). The consistency is then applied between the mappings of the anchor image which now lie on different parts of the circuit.  This additional measure ensures robustness by making the entire output of the circuit comply with the decided-upon separability as opposed to just a few qubits.  This technique also enables scaling to entanglement of more than 2 images on a single machine. 

In Fig. \ref{fig:sliq}, we show an overview for the design of \textsc{SliQ}. The first step is to take the input dataset and create pairs of the form (Anchor, Positive) and (Anchor, Negative) for training and testing.  Once these are formed, the network trains on the dataset classically as part of the hybrid quantum-classical model used in VQCs. Once the data is trained, \textsc{SliQ} performs similarity detection by performing inference on new pairs of data. This inference can be used to identify the most similar samples to one another, the most distant samples, and even serve as a classifier if an unsupervised clustering algorithm is used to generate clusters.

\section{Experimental Methodology}
\label{sec:metho}

\noindent\textbf{Training and Testing Datasets.} \textsc{SliQ} is evaluated on NIH AIDS Antiviral Screen Data~\cite{kramer2001molecular}, MNIST~\cite{mnist}, Fashion-MNIST~\cite{xiao2017/online}, and Flickr Landscape \cite{chen2018cartoongan}. These datasets are chosen because (1) they cover both unlabeled (e.g., Flickr Landscape) and labeled datasets (e.g., AIDS, MNIST, Fashion-MNIST), (2) they represent different characteristics (e.g., medical dataset, images, handwritten characters) and have real-world use cases (e.g., AIDS detection). We note that the size of the Flickr Landscape dataset is $\approx$25$\times$ larger than the commonly used quantum machine learning image datasets of MNIST and Fashion-MNIST~\cite{xiao2017/online,silver2022quilt}. This presents additional scaling challenges that we mitigate with the scalable design of \textsc{SliQ}.

Flickr Landscape is an unlabeled dataset consisting of 4300 images of different landscapes spanning general landscapes, mountains, deserts, seas, beaches, islands, and Japan. The images are different sizes, but are cropped to 80$\times$80$\times$3 for consistency.  The center is kept intact with all color channels. This dataset is unlabeled and serves to show how \textsc{SliQ} performs on unlabeled data. NIH AIDS Antiviral Screen Data contains 50,000 samples of features alongside a label to indicate the status of the Aids virus (``CA'' for confirmed active, ``CM'' for confirmed moderately active, and ``CI'' for confirmed inactive). The MNIST dataset is a grayscale handwritten digit dataset~\cite{mnist} where we perform binary classification on `1's and `0's. The Fashion-MNIST dataset contains the same number of pixels and classes as MNIST, but the images are more complex in nature.  In all datasets, 80\% of the data is reserved for training and 20\% reserved for testing.

\noindent\textbf{Experimental Framework.}The environment for \textsc{SliQ} is Python3 with Pennylane \cite{pennylane} and Qiskit \cite{Qiskit} frameworks. Our quantum experiment evaluations are simulated classically, and the inference results which are collected on the real IBM quantum machines are specified in the text. For setting up the datasets for testing and training, triplets are created in the form of an anchor image, a positive image, and a negative image. 

For the unlabeled dataset used, three images are selected at random.  The color frequency is evaluated for each image on R, G, and B channels individually, then placed into eight bins for each color. The 24 total bins for each image are used to establish a ground truth similarity between the images, where the first image is the anchor and the images closest in L1 norm to the anchor is set as the positive image and the further away image is the negative image.  Once the triplets are created, the pixels within the images are interwoven with one another.  The image is then padded with 0s to the nearest power of 2 for the amplitude embedding process.  

In the case of the labeled datasets, the anchor and positive are chosen to be from the same class, while the negative image is chosen at random from another class. For evaluation, an anchor and positive are passed into the embedding and the four dimensions are used in a Gaussian Mixture Model to form clusters which are then evaluated for accuracy. We use a batch size of 30 for all experiments with a GradientDescentOptimizer and a learning rate of $0.01$.  We train for 500 epochs on a four-layer network. The size of the network changes based on the dataset used, a four-qubit circuit is used for the Aids dataset, 14 for Flickr Landscape, and 11 for MNIST and Fashion-MNIST. For the baseline, one less qubit is used for all circuits as it does not necessitate the additional qubit to store an additional sample on each run.

\noindent\textbf{Competing Techniques.} Our baseline scheme is the same as described earlier: it is a quantum analogue of a triplet network, where weights are shared and use a single network for training and inference. Although \textsc{SliQ} is not designed for classification tasks on labeled data, we provide a quantitative comparison with also the state-of-the-art quantum machine learning classifier: Projected Quantum Kernel (PQK) (published in Nature'2021)~\cite{49725} and Quilt (published in AAAI'2022)~\cite{silver2022quilt}. The former, PQK, trains a classical network on quantum features generated by a specially designed quantum circuit. Datasets are modified to have quantum features that show the quantum advantage in training. While not feasible to implement on current quantum computers, we modify PQK's architecture to use fewer intermediate layers to reduce the number of gates and training time. The other comparison used, Quilt, is an image classifier built on an ensemble of variational quantum circuits which uses the ensemble to mitigate noise in the NISQ-era of quantum computing.

\noindent\textbf{Figures of Merit.} We categorize the figures of merit by dataset type: unlabeled or labeled. As image similarity is inherently visual, in addition to quantitative success, we demonstrate qualitative success by including a few relevant snapshots. For our unlabeled results, we also examine quantitatively how well the overall predictions match with the ground truth similarities, showing how well \textsc{SliQ} learns over a distribution. The specific details of the metrics are described near the description of the results. For qualitative success in the unlabeled dataset, we show the model's predicted images for most similar to a given anchor image. For our labeled datasets, we report the accuracy of \textsc{SliQ}. Because \textsc{SliQ} performs an embedding and does not classify, accuracy can not be obtained directly from the output.  For accuracy metrics, Gaussian Mixture Models are used to assign clusters for classification.

\section{Evaluation and Analysis}
\label{sec:evalu}

\begin{table}[t]
\begin{center}
\label{tab:distribution}
\scalebox{0.86}{
\begin{tabular}{|c|cc||c|cc|} 
 \hline
 Percentile & Baseline & \textsc{SliQ} & Percentile & Baseline & \textsc{SliQ} \\ [0.5ex] 
 \hline\hline
  25$^{th}$ & \( -0.06\) & \( 0.23\) & 75$^{th}$ & \( 0.19\) & \( 0.48\) \\ 
  \hline
  50$^{th}$ & \( 0.05\) & \( 0.36\) & 100$^{th}$ & \( 0.63\) & \( 0.68\) \\ 
 \hline
\end{tabular}}
\end{center}
\caption{Spearman correlation results show that \textsc{SliQ} outperforms the baseline in similarity detection. \textsc{SliQ}'s median Spearman correlation is 0.36 vs. the Baseline's 0.05.}

\end{table}

\begin{figure}[t]
    \centering
    \includegraphics[scale=0.50]{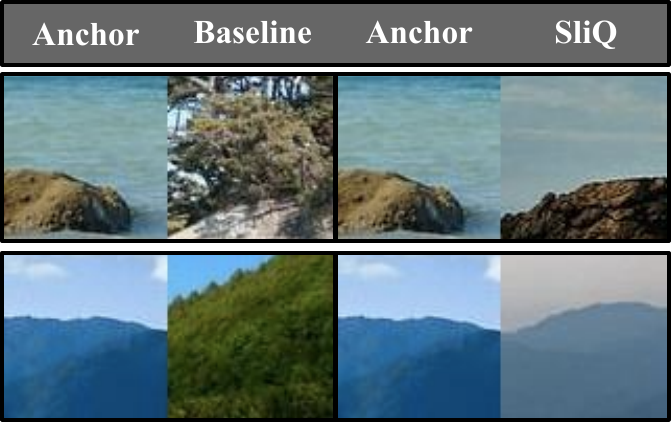}
    \caption{Anchor image, Baseline-identified similar image, and \textsc{SliQ}-identified similar image for the same anchor image, using the unlabeled Flickr Dataset. \textsc{SliQ} is effective in identifying images with similar color frequencies.}
    \label{closest}
\end{figure}

\noindent\textbf{\textsc{SliQ} effectively detects the similarity of samples in unlabeled datasets using quantum networks; \textsc{SliQ} is the first work to target this area.} As \textsc{SliQ} is the first work to target quantum similarity detection for unlabeled data, we compare against the results for the baseline quantum triplet model. Using the Flickr Landscape dataset, we rank the image similarity based on triplets formed from color frequency histograms. For each image in a set of 100 images, we treat the image as an anchor and compute the ground truth distance in color frequency between the anchor and every other image. We compare this ground truth distance to the distance identified by the model and correlate the rankings. 

We use Spearman correlation \cite{myers2004spearman}, which performs ranked correlation of two random variables, to interpret the relationship between ground truth and the model's estimations. Spearman correlation is commonly used to perform this type of analysis, for example \cite{reimers2019sentence} uses Spearman correlation in ranking sentence embedding in similarity networks. \textit{\textsc{SliQ} has much better correlations over the baseline triplet model, with a median Spearman correlation of 0.36 compared to a median Spearman correlation of 0.05, which shows that \textsc{SliQ} is 0.31 or 31\% points more accurately correlated than Baseline.} In Table \ref{tab:distribution}, we show the distribution of Spearman correlations for \textsc{SliQ} compared to the baselines. At every percentile measured, \textsc{SliQ} has notable improvements in similarity detection, which demonstrates \textsc{SliQ}'s overall improvement over an entire distribution. 

This trend is also visually observed in Fig.~\ref{ranking}(a) for Baseline and Fig.~\ref{ranking}(b) for \textsc{SliQ}. The x-axis denotes the ground truth distance, where the points further to the left indicate true similarity. The y-axis denotes the calculated loss of the model, indicating \textsc{SliQ}'s ground truth distance estimate. Points closer to the diagonal line denote accurate estimations. Fig.~\ref{ranking} show only the top-3 correlation examples for easier visual interpretation. We note that \textsc{SliQ} tends to cluster more around the ground truth correlation line, and its top-3 correlation drop from 0.68 to 0.64; in contrast, the baseline, drops from 0.63 to 0.46.

Additionally, we show some of the closest predictions identified in Fig.~\ref{closest}. These demonstrate the different scenes of the Landscape dataset.  For example, the demonstrated landscapes include mountains, aerial perspectives, forests, and oceans. \textsc{SliQ} is able to match similar landscapes efficiently, demonstrating its effectiveness at similarity detection -- better than the baseline's relative picks.

\begin{figure}[t]
    \centering
    \includegraphics[scale=0.50]{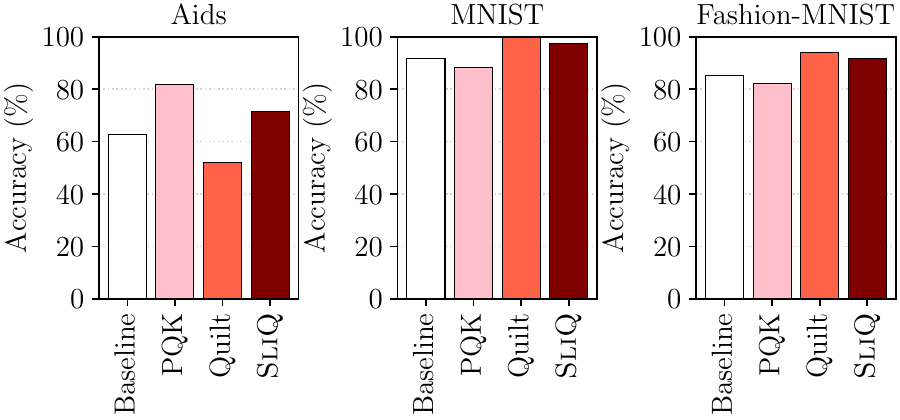}
    \caption{\textsc{SliQ} performs competitively against the comparative classification techniques for all datasets, despite classification not being a learned objective of \textsc{SliQ}.}
    \label{fig:main}
\end{figure}

\begin{table}[t]
\begin{center}
\label{tab:real}
\begin{tabular}{|c|cccc|} 
 \hline
 Environment & Baseline & PQK & Quilt & \textsc{SliQ} \\ [0.5ex] 
 \hline\hline
 Simulation & \ 60.8\%\ & 81.6\% & 51\%\ &  71.54\%\ \\ 
  \hline
 Real Computer & \ 64.4\%\ & N/A & 21.4\%\ & \ 68.8\%\ \\ 
 \hline
\end{tabular}
\end{center}
\caption{\textsc{SliQ}'s real quantum computer results for the AIDS dataset are consistent with the simulation results, showing its resilient to hardware noise due to its low number of parameters. The PQK column is denoted as N/A circuit is prohibitively deep for it to be feasible and run on error-prone NISQ computers (30$\times$ more parameters than \textsc{SliQ}).}

\end{table}

\noindent\textbf{Although \textsc{SliQ} was not designed to act as a classifier, it is effective at detecting the similarity of samples in labeled datasets and is competitive with prior state-of-the-art quantum classifiers.} On its own, \textsc{SliQ} performs embeddings, not classification, but we can use \textsc{SliQ} as a classifier by performing reasonable clustering on its output. To demonstrate its classification ability, we compare against state-of the art quantum classification methods \cite{49725,silver2022quilt}. Our results (Fig.~\ref{fig:main}) indicate that \textsc{SliQ} yields competitive accuracy compared to the advanced quantum classifiers on a task that \textsc{SliQ} was never trained on (classification) and demonstrates the broad applicability of embeddings. To perform this clustering, we employ a Gaussian Mixture Model on our embeddings.  The model is initialized with the true number of classes in the dataset and is fit to 1000 samples to form clusters.  For classification, each permutation of the clusters to labels is considered, as these models do not provide labels. The permutation with the highest accuracy is considered to be the correct label. With these techniques, our results show \textsc{SliQ} performs well on a variety of datasets, averaging up to 96.44\% accuracy on binary MNIST classification. We show the full classification accuracy results in Fig. \ref{fig:main} for different techniques.

In Table~\ref{tab:real}, we show how \textsc{SliQ} performs on real quantum computers today. \textsc{SliQ} achieves a 68.8\% accuracy on the AIDS dataset, running on IBM Oslo. \textit{\textsc{SliQ} significantly outperforms the state-of-the-art quantum classifier (QUILT), even though \textsc{SliQ} was not originally designed for classification.} The reason is because \textsc{SliQ}'s design is noise-aware to be effective on error-prone NISQ computers. In particular, \textsc{SliQ} is designed with few parameters for current NISQ computers, where error compounds at each step and quickly explodes.  Other quantum machine learning architectures tend to have higher degradation in accuracy on real computers as they require larger architectures with ample opportunity for compounding error. For the AIDS dataset, \textsc{SliQ} has 48 tunable parameters, while Quilt requires 375, and PQK requires \numprint{1633} parameters. As a result of more parameters, the hardware error compounds, explaining the degradation shown above.

\begin{figure}[t]
    \centering
    \includegraphics[scale=0.55]{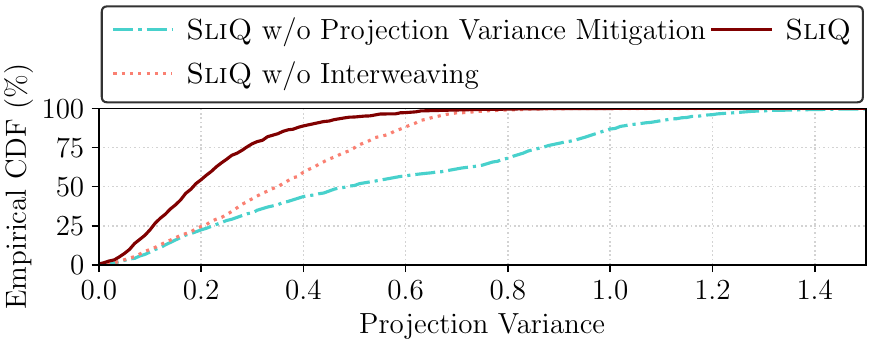}
    \caption{\textsc{SliQ} achieves a lower projection variance compared to when it is run without projection variance mitigation or without interweaving of images.}
    \label{invariance}
\end{figure}

\noindent\textbf{Why does \textsc{SliQ} perform effectively?} By mitigating projection variance, \textsc{SliQ} is able to map the anchor input at the same position to a close location regardless of the second input in the pair.  This is necessary as the images get entangled together throughout the entire circuit and will not be separated in the output unless a constraint is put in place to enforce this. This separability can be demonstrated in Fig.~\ref{invariance} where \textsc{SliQ} is compared to a trained version without consistency loss. \textsc{SliQ} has more precise outputs throughout the entire CDF, evaluated over 1000 MNIST test images. Enforcing consistency amounts to an average decrease of 80\% in projection variance when changing the order of the inputs -- demonstrating the effectiveness of \textsc{SliQ}'s projection invariance method.
As shown in Fig. \ref{invariance}, interweaving increasing robustness, leading to a decrease in projection variance. 

\section{Related Work}
\label{sec:relat}

\noindent\textbf{Classical Similarity Networks.} Siamese networks and triplet networks are commonly-used classical similarity networks~\cite{8733051,Koch2015SiameseNN,7298682,9010055,10.1007/978-3-030-58621-8_9,designs4020009,10.1093/bib/bbaa266,hoffer2015deep,https://doi.org/10.48550/arxiv.2205.04051}, as they known are to be the best choice for complex datasets~\cite{Chicco2021}. As an instance, using the Riemannian geometry to train the Siamese network, Roy et al.~\cite{9010055} get effective results for image classification, while FaceNet~\cite{7298682} achieves representational efficiency using an embedding for face recognition and clustering. On the other hand, TrimNet~\cite{10.1093/bib/bbaa266} uses a graph-based approach toward enabling a triplet network to learn molecular representations. However, while these works are effective classically, quantum theory enables the enhancement of machine learning workloads by reducing their size and speeding them up~\cite{Aaronson2015,Daley2022}.

\noindent\textbf{Quantum Machine Learning.} Extensive work has been performed toward porting a wide variety of machine learning tasks to quantum computing~\cite{50941,Li2021,Tiwari_Melucci_2019,Li_Song_Wang_2021,Khairy_Shaydulin_Cincio_Alexeev_Balaprakash_2020,Lockwood_Si_2020,Heidari_Grama_Szpankowski_2022,https://doi.org/10.48550/arxiv.2001.03622,https://doi.org/10.48550/arxiv.2201.02310,https://doi.org/10.1049/qtc2.12026}. This includes porting workloads such as generalized neural networks~\cite{Beer2020}, convolutional neural networks~\cite{Hur2022}, and even application-specific networks such as models used to learn the metal-insulator transition of VO$_{2}$~\cite{Li2021}.

\noindent\textbf{Quantum Image Similarity Detection} 
\cite{liu2022comparison} and \cite{liu2019similarity} have also worked on quantum image similarity detection; notably, these did not take a machine-learning approach to identify similarity.

\section{Conclusion}
\label{sec:concl}

In this work, we present \textsc{SliQ}, a resource-efficient quantum similarity network, which is the first method to build a practical and effective quantum learning circuit for similarity detection on NISQ-era computers. We show that \textsc{SliQ} improves similarity detection over a baseline quantum triplet network by 31\% points for Spearman correlation. \textsc{SliQ} is available at: \texttt{https://github.com/SilverEngineered/SliQ}.

\section{Acknowledgements} We thank the anonymous reviewers for their constructive feedback. This work was supported in part by Northeastern University and NSF Award 2144540. 

\bibliography{aaai23}

\end{document}